\documentclass[journal]{IEEEtran}

\usepackage{epsfig}

\newcommand{\n}{n}                  
\newcommand{\p}{p}                  
\newcommand{\up}{b}                 
\newcommand{\down}{a}               
\newcommand{\rGR}{ \GR }            
\newcommand{\GR}{ \mu_{\n} }        
\newcommand{\gr}{ \mu_{\p} }        
\newcommand{\CR}{ R }               
\newcommand{\Ecoli}{\emph{E. coli}}

\makeatletter
\renewcommand\@biblabel[1]{#1.}
\makeatother

\begin{document}

\title{Multi-Objective Optimization Applied
to the Eradication of Persistent Pathogens}

\author{Ole Steuernagel,$^{1}$ Daniel Polani,$^{2}$%
\thanks{O.Steuernagel@herts.ac.uk, D.Polani@herts.ac.uk}
\thanks{${}^1$School of Physics, Astronomy and Mathematics,
University of Hertfordshire, Hatfield, AL10 9AB, UK}
\thanks{${}^2$School of Computer Science, University of Hertfordshire,
Hatfield, AL10 9AB, UK}
}


\maketitle

\begin{abstract}

  In scenarios such as therapeutic modelling or pest control, one aims
  to suppress infective agents or maximize crop yields while
  minimizing the side-effects of interventions, such as cost,
  environmental impact, and toxicity.  Here, we consider the
  eradication of persistent microbes (e.g. \Ecoli, \emph{Multiply
    Resistent Staphylococcus aureus} (MRSA-`superbug'),
  \emph{Mycobacterium tuberculosis}, \emph{Pseudomonas aeruginosa})
  through medication. Such microbe populations consist of
  metabolically active and metabolically inactive (persistent)
  subpopulations. It turns out that, for efficient medication
  strategies, the two goals, eradication of active bacteria on one
  hand and eradication of inactive bacteria on the other, are in
  conflict.  Using multi-objective optimization, we obtain a survey of
  the full spectrum of best solutions.  We find that, if treatment
  time is limited and the total medication dose is constant, the
  application of the medication should be concentrated both at the
  beginning and end of the treatment.  If the treatment time is
  increased, the medication should become increasingly spread out over
  the treatment period until it is uniformly spread over the entire
  period. The transition between short and long overall treatment
  times sees optimal medication strategies clustered into groups.

\end{abstract}



\IEEEpeerreviewmaketitle

\section*{Introduction}

\subsection*{The problem of bacterial persistence}

Not only hibernating mammals or sporing fungi reduce or stop their
metabolic activities, also some microbial organisms are known to
randomly slip into and out of `hibernation': this is essentially
characterized by reduced metabolic activity and reduced or suspended
reproduction. The disadvantage of reduced population growth, goes
hand-in-hand with the advantage of reduced vulnerability to drugs,
rendering `hibernating' bacteria persistent in the face of
medication
treatments~\cite{Balaban04,Levin04,Keren04,Finzi99,WiuffAAC05}.
Bacterial population can therefore consist of genetically identical
active and persister subpopulations.

From a human point of view, be it medical or pest control, the
presence of persisters can have serious consequences. Bacterial
persistence was first observed in \emph{Staphylococcus} when in 1944
Bigger~\cite{Balaban04,Keren04} noted that penicillin did not always
kill all exposed bacteria although sufficient toxicity was
established. Based on the observation that bacteria surviving
penicillin treatment were no less susceptible than their ancestors,
it was concluded that heritable bacterial resistance was not
involved but persistent behavi\-our could explain such a finding;
this has recently been re-confirmed~\cite{Balaban04}.

Bacterial persistence can occur irrespective of environmental
conditions~\cite{Balaban04,Levin04} and is
widespread~\cite{Balaban04,Levin04,Keren04,Finzi99,Hernandez99,delGiorgio97,Lebaron01}.
It also appears in viruses, which can become persistent by
integrating into their host's genome and suspending production of
virus particles, as exemplified by HIV, herpes, and the
bacteriophage lambda.

The tradeoff between the persisters' growth-underperformance under
benign conditions on one hand, and the wipeout of all active
organisms in the case of a catastrophe on the other, leads to a
small random subpopulation of persisters individually
``bet-hedging'' to switch into a persistent state~\cite{Gardner07}
thus effectively establishing a
``life-insurance''~\cite{KusselGenetics05,KusselSCI05}. Persistence
can bring a species ``back from the brink'', even when sequences of
sudden catastrophes occur, because, in all likelihood, a few
persisters will have stayed out of harm's way~\cite{KusselSCI05}. It
is thus relevant in disease prevention~\cite{Keren04} and requires
new treatment regimes~\cite{WiuffAAC05}.  \Ecoli,
\emph{MRSA-`superbug'}, \emph{Mycobacterium tuberculosis}, and
\emph{Pseudomonas aeruginosa} show
persistence~\cite{Balaban04,Levin04,Keren04}, possibly extending to
airborne infectants~\cite{Hernandez99}. Persistence also appears to
be important in ecological scenarios~\cite{delGiorgio97,Lebaron01}
and latent \emph{HIV-1} infections~\cite{Finzi99}.

In \Ecoli,~the conversion rate from the active subpopulation to the persistent
form and the reverse rate have been shown to be independent of environmental
factors~\cite{Balaban04,Keren04}. In other words, no sensorial input about the
quality of the environment is used to trigger the conversion from one to the
other: here we only consider this type of persistence. It is an effective
strategy for organisms which face life in environments where sudden
de\-vas\-ta\-ting degradation and recovery is an acute possibility and
moreover dispenses with the need to maintain sensors for surveying the
environment -- an important advantage for primitive
organisms~\cite{KusselSCI05}.
\subsection*{Our approach}
In this paper we primarily intend to highlight the features of multi-objective
optimization and its applicability to problems in the life sciences. Our model
for the eradication of persistent pathogens shows that eradication of
persisters and normal pathogens form conflicting objectives which are best
approached using multi-objective optimization. Our model is not intended to
quantitatively represent a specific biological or clinical system but to
investigate the problem in general terms. Multi-objective
optimization~\cite{deb99Algorithm} has been
rarely applied to problems in the life sciences~\cite{Lahanas03,Handl07}.
It is, however, becoming clear that the benefits of using multi-objective
optimization in the life sciences could be considerable~\cite{Handl07}.

We adopt popula\-tion dynamical models based on coupled differential
equations~\cite{Balaban04,Levin04,KusselGenetics05,KusselSCI05} for the
numerical study of the behaviour of persistent pathogens exposed to different
medication strategies (we use the terms `medication' and `drug'
interchangeably, they stand for the presence of any hazardous entity killing
the pathogens, such as radiation, chemicals, antibodies, etc.).

We assume that only two subpopulations are present, active (normal) pathogens
that grow at a normal rate, and are susceptible to the medication, and
persisters (such as type~II in reference~\cite{Balaban04}) that grow more
slowly and are less susceptible. We assume that the subpopulations are so large
that discreteness of population sizes can be neglected. we denote the sizes of
the normal and the persistent subpopulations by the time-dependent functions,
$\n(t)$ and $\p(t)$, respectively. This continuous description allows us to
employ continuous differential equations (in the `deterministic
limit'~\cite{KusselGenetics05}) which are readily integrated using a computer.

Initially, we will confirm mathematically that the best approach to the
eradication of non-persistent multiplying pathogens is their immediate
extermination by as strong a medication dose as possible. Whereas this case is
intuitively easy to understand, matters become much more complicated when
persistence is taken into account.

The slowdown or shutdown of the persisters' metabolism protects them from
medication. One therefore has to retain some medication to be administered some
time after the first dose of medication was applied. This helps to exterminate
persistent pathogens that have bypassed the biocidal effects of the initially
given medication and subsequently revert back to their active state.

For such followup action neither very long waiting times are allowed, because
the surviving active pathogens multiply and thus hurt the host and replenish
the persisters' reservoir, nor is immediate followup medication advised;
otherwise the persisters have not had enough time to come out of the persistent
state and so the medication hurts the host more than the pathogens.

Neither intuitive nor analytical solutions for this problem are available, we
therefore choose a model in which a course of treatment consists of the
administration of $N$ equal units of the drug (we choose $N=10$). The total
amount of drug applied during a course of treatment is fixed. The course of
treatment extends over a fixed time interval, spanning from the initial time
$t=0$, of the pathogens' detection, to the time $t=T$ when the final outcome
of the treatment is evaluated. Within this interval, times for the individual
drug administrations~$t_k$, with ($k=1,...,N$), are chosen freely. Different
medication scenarios, i.e. the effect of different distributions of the $N$
administration times $\{ t_k \}$ are compared for their effectiveness. The
objective of the treatment is the minimization of the sizes of the
normal,~$\n(T)$, and the persistent subpopulation,~$\p(T)$, at the end of the
course of treatment.

Although we perform multi-objective optimization we fix this treatment time,
$T$, beforehand. One can, of course, generalize our approach to include
variable treatment times as well, thus having to consider three objective
variables, namely, $\n(T)$, $\p(T)$, and $T$. Then, our problem space would be
three dimensional and the set of best solutions would form a complicated
two-dimensional hypersurface embedded in it: too rich a system for an
introductory treatment of our method. We thus consider a two-dimensional
problem space $[n(T),p(T)]$ and the family of best solutions that form a
one-dimensional hypersurface within.

We will, towards the end of this paper consider the general trends of our model
system's behaviour when the total treatment time $T$ is varied as well.

Aside from this simplicity issue, there are two more good reasons to fix the
total treatment time $T$ beforehand.

We assume that cumulative toxicity of the medication is the major constraint
regarding its application (this is reasonable for scenarios, such as
radiation therapy, many types of medication treatments and for agricultural
and other such environmental scenarios). With a cumulative dosage constraint,
medication strategies must not be drawn out too much in time since the
medication becomes overdiluted, see below. We therefore arrive at a natural
upper limit for the total treatment time~$T$.

If there is no time constraint, and if one makes sure that the
medication does not become overdiluted, drawn out medication regimes
where the medication is administered at a constant rate throughout the
treatment $T$ show the greatest suppression of~$\n(T)$ and $\p(T)$,
see Figure~\ref{fig3} below. But this kind of treatment regime can
become unstable due to the danger of overdilution (see caption of
Figure~\ref{fig3} below) and is also harder to adhere to than
treatment regimes of fixed shorter time.

In light of the fact that imperfect patient adherence to medication strategies
is of considerable concern~\cite{Atreja05}, we thus conclude that there are
the following reasons to consider fixed total treatment times~$T$: simplicity,
safety, and practicality.

\subsection*{The problem space and the Pareto front}

Different medication strategies lead to different final results. When the
times at which a medication dose is administered is continuously changed the
outcome changes continuously as well. Therefore, the problem space consists of
a connected area of feasible solutions outside of which lies the region which
cannot be reached by any feasible solution; because, say, perfect or near
perfect suppression of the pathogens' numbers is beyond the eradication power
of the medication. An example in our model would be the origin
$[\n(T),\p(T)]=[0,0]$ and its immediate neighbourhood. This area cannot be
reached because the differential equations used in our model only allow for
exponential suppression of the population, not complete eradication (on the
issue of complete extermination due to fluctuations see
reference~\cite{KusselGenetics05}).

The most interesting area is the boundary that lies between feasible and
unfeasible solutions for small numbers of $\n(T)$ and $\p(T)$, because it
contains the optimal cases of what is feasible. The boundary can have a
complicated shape, see, for instance, Figs.~10 and~13 of
reference~\cite{Messac03}.

The \emph{Pareto front} contains all those points of the boundary for which
there are no other points which allow for solutions that are {\em
simultaneously} better or equal with respect to all optimization objectives; it
only contains `non-dominated' solutions. Since the boundary can have a
complicated form, the subset of Pareto-optimal points can be discontinuous, see
Figs.~11~(c) and~14~(b) of reference~\cite{Messac03}. Typically, a continuous
Pareto front in two dimensions has a shape like the curve shown in
Fig.~\ref{fig2}~\textbf{b} below, also, compare Figs.~9,~10 and~13 in
reference~\cite{Messac03}.

In principle, using conventional single-objective optimization and changing all
available relative weight factors that were used to combine several objectives
artificially into a single one allows us to find the Pareto front as
well~\cite{Messac03}. However, for practical reasons this modification of
single-objective optimization is unfeasible because many solutions are being
entirely missed, see Fig.~14~(c) of reference~\cite{Messac03}. Using
multi-objective optimization allows us to gain the advantage of being able to
explore the entire set of optimal solutions~\cite{Messac03}.

\subsection*{Our model}

For transparency we also employ the following simplifications:

The active subpopulation,~$\n(t) $, grows at a constant rate~$\GR$
leading to exponential growth, whereas the persisting
subpopulation,~$\p(t) $, grows at a substantially lower
rate~$\gr$~\cite{KusselGenetics05} which we set to zero for simplicity
(without affecting our basic conclusions). We, similarly, neglect the
(greatly reduced) kill rate of persisters in the presence of
medication~\cite{Balaban04,KusselGenetics05}.

The subpopulations convert into each other at constant rates $\down$
and $\up$~\cite{Balaban04,KusselGenetics05}, although these rates
may depend on environmental conditions~\cite{Gardner07}. We assume
that only the active subpopulation is being decimated by the
medication: we assume its power to kill to be proportional to the
drug concentration,~$c(t)$,~\cite{Boonkitticharoen90} (although
non-linear threshold behaviour has been observed as
well~\cite{Boonkitticharoen90} -- in which case other assumptions
such as zero growth of the persisters may have to be reviewed). We
therefore arrive at the following system of coupled ordinary
differential equations for the behaviour of the subpopulations as
functions of time
\begin{eqnarray}
\label{original_1} \frac{d \, \n(t)}{dt} & = & ( \GR -c(t)-\down)
\cdot \n(t) + \up \cdot \p(t) \; ,
\\
\label{original_2}
\frac{d \, \p(t)}{dt} & = &\down\cdot \n(t)   - \up \cdot \p(t)\;
.
\end{eqnarray}

\subsection*{Our assumptions}

In what follows we will assume that the total administered medication dose is
fixed. This assumption is motivated by the cumulative toxicity of medical
treatments. Our approach can be adapted accordingly, if avoidance of peak
values of the drug concentration is the primary concern.

In general the concentration of the drug,~$c(t)$, could be given by
any nonnegative function. In accord with our approach we model each
administered drug dose by the same Gaussian peaks (bell-shaped curves)
with equal strength~$D_0$, centered on the respective administration
times~$t_k$, a treatment course, $D(t)$, is thus described by the sum
\begin{eqnarray}
\label{Dosages.over.time}
D(t)=\sum_{k=1}^N D_0 \; \frac{\exp\left[{
\frac{-(t-t_k)^2}{\sigma^2}}\right]}{\sqrt{\pi} \sigma} \; .
\end{eqnarray}
Here $\sigma$ scales the widths of the Gaussians (compare Fig.~\ref{fig1}) and
the normalization factor $1/(\sqrt{\pi}\sigma)$ assures that each peak is of
unit strength ($\int_{-\infty}^{\infty} dt \exp
\left[\frac{-(t-t_k)^2}{\sigma^2}\right]/({\sqrt{\pi} \sigma}) = 1$).

We assume that the drug is cleared out of the system at a constant rate~$\CR$
(in units of hr${}^{-1}$), its concentration,~$c(t)$, thus obeys the
differential equation
\begin{eqnarray}
\label{c.dgl} \frac{d \, c(t)}{dt} & = & D(t) - \CR \; c(t) \; .
\end{eqnarray}
For the drug concentration this yields
\begin{eqnarray}
\label{c.sol} c(t) & = & c_0 + \int_{0}^t d\tau \; D(\tau) \;
e^{-\CR (t-\tau)}  \; ,
\end{eqnarray}
with the assumed initial value $c_0=0$, i.e. no medication is present before
the treatment starts at time zero. Note that small values of the drug-clearance
rate $\CR$ lead to prolonged presence of the medication and thus to a greater
cumulative effect since the accumulated medication dose
\begin{eqnarray}
\label{C.Integrated} C(T) = \int_{0}^T dt \; c(t)
\label{integrated.medication}
\end{eqnarray}
scales with $R^{-1}$, just like the total integrated dose
\begin{equation}
C_\infty \doteq \int_{-\infty}^\infty \, dt \, \int_{-\infty}^t \, d\tau \,
D(\tau) e^{-\CR (t-\tau)} = N \, D_0\, / \, R \; . \label{integral.medication}
\end{equation}
This expression $C_\infty$ ignores the initial value assumption
$c_0=0$ and therefore has a simple and transparent form. Because it
includes the small tails of the medication distribution extending
beyond of the treatment time interval~$[0,T]$ it slightly
overestimates the value of the total cumulative administered
dose,~$C(T)$.

Scenarios that can be described by eqs.~(\ref{original_1})
and~(\ref{original_2}) include a bacterial infection by a persistent species
which is being fought with drugs (term `$-c(t) \cdot n(t)$' in
eq.~(\ref{c.dgl})) where the drug degrades over time (say, by excretion,
aging, or evaporation: term `$e^{-\CR (t-\tau)}$' in
eq.~(\ref{integral.medication})).

Note that our assumption regarding a finite number of administered
doses $N$, in eq.~(\ref{Dosages.over.time}), is the generic way in
which medication is released (in radiation treatment or pest control
with agricultural aircraft, continuous administration may be
altogether unfeasible). Continuous medication (drip-feed) can easily
be emulated with our model using a large number, $N$, of doses shots.

Without medication ($c(t)=0$) the system~(\ref{original_1})-(\ref{original_2})
has constant coefficients and is therefore analytically solvable with the
general solution $[n(t),p(t)] = w_{+} \vec{e}_+ \exp[\lambda_+ t] + w_{-}
\vec{e}_- \exp[\lambda_- t]$. Here, the two eigenvector $\vec{e}_\pm =
[\GR-\down+\up \pm \sqrt{(\down+\up)^2+\GR(
 \GR -2 \down + 2 \up)}, 2\down]$
are associated with eigenvalues $\lambda_\pm = [\up \rGR \pm \sqrt{\up^2 \rGR^2
+ 4 ( \rGR - \up - \down )}]/2 $. The component $w_{+} \vec{e}_+$ will quickly
outgrow its counterpart $w_{-} \vec{e}_-$ because $\lambda_+ > \lambda_-$. The
wild-type without the influence of medication is therefore typically well
described by the state $w_{+} \vec{e}_+ \exp[\lambda_+ t]$. This implies that
the generic initial ratio of active to persistent pathogens is given by the
ratio of the components of the wild-type $\vec e_+$, namely
\begin{eqnarray}
\label{init_ratio}
\frac{{\n}(0)}{{\p}(0)}
= \frac{\GR +\up - \down +\sqrt{(\up+\down)^2+2\GR(\up-\down)+\GR^2} }{2\down}
.
\end{eqnarray}
We therefore choose $\n(0)=1$ and $\p(0)$ in accordance with
eq.~(\ref{init_ratio}) as a \emph{natural initial} condition which models
pathogens found in their natural infection habitat.  If they are found under
very different circumstances, such as bacteria residing in a bacterial biofilm,
the initial fraction of persisters can be much higher~\cite{Keren04} than
assumed here.

\section*{Results}
\begin{figure}
\begin{center}

\hspace{1.5cm} \textbf{a}\\

\hspace{2cm}
\includegraphics[width=3.2in,height=2in]{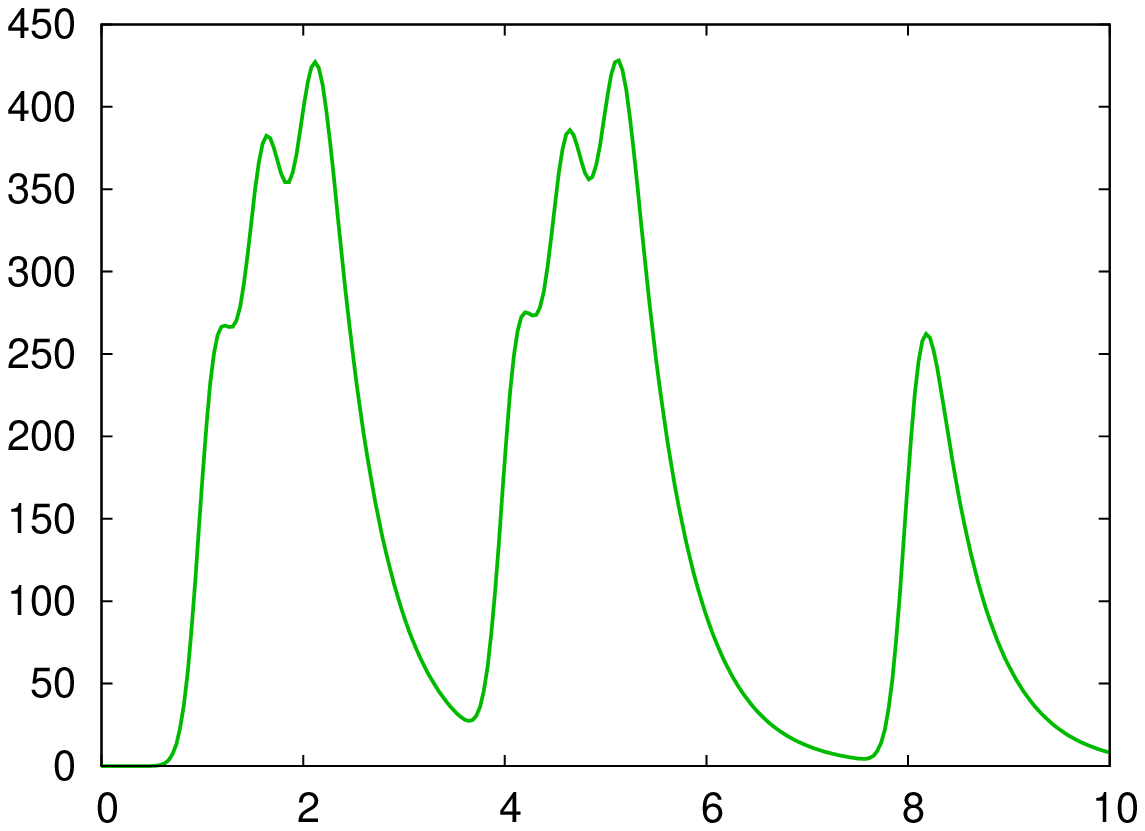}
\put(-240,20){\rotatebox{90}{\mbox{Medication Concentration}}}
\put(-30,-15){Time (h)}
\\

\vspace{1cm} \hspace{1.5cm} \textbf{b}\\

\hspace{1.5cm}
\includegraphics[width=3.2in,height=2in]{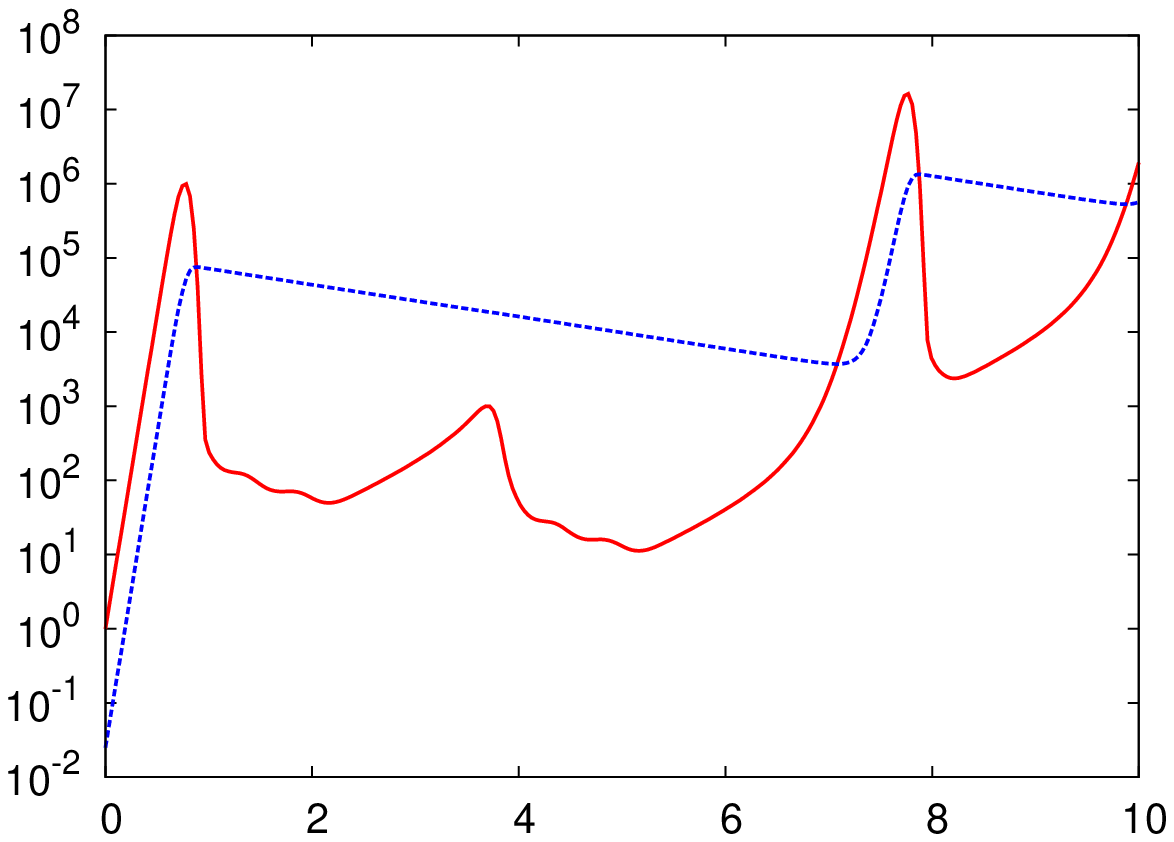}
\put(-240,30){\rotatebox{90}{\mbox{Pathogen Population}}}
\put(-30,-10){Time (h)}
\end{center}

\caption{\sf \small Response of pathogen population to
  medication. {\bf a}, drug concentration, $c(t)$, as a function of
  time. $N=7$ dose units, of strength $D_0=430$ each, are administered
  at times $1, 1.5, 2, 4, 4.5, 5, 8$ (hours). The widths of the peaks
  is $\sigma=0.2 \,\mathrm{h}$ and the drug-clearance rate is
  $\CR=2\,\mathrm{h}^{-1}$. {\bf b}, The time evolution of active
  organisms $\n(t)$ (red solid line) and persisters $\p(t)$ (blue
  dotted line) with parameters $\GR=20\,\mathrm{h}^{-1}$,
  $\down=0.5\,\mathrm{h}^{-1}$ and $\up=0.5\,\mathrm{h}^{-1}$, note
  that the persister responses are delayed in time.}
 \label{fig1}
\end{figure}

Equations~(\ref{original_1}) and~(\ref{original_2}) do not, in general, allow
for an analytical solution. This is why we investigate them numerically. A
typical scenario is portrayed in Figure~\ref{fig1}. It illustrates that active
organisms, $\n (t)$, primarily get killed by the medication whereas the
inactive ones, $\p (t)$, primarily suffer losses due to conversion into active
ones and regain numbers when the active ones recover. The two subpopulations
sustain each other.

The goal is to push the entire pathogen population towards its possible
extinction (i.e. to such small numbers that action of the host's immune system
or random fluctuations can wipe it out~\cite{KusselGenetics05}).

\subsection*{Fighting non-persistent pathogens}

Without persistence all pathogens are affected by the medication and
should be killed immediately. This can be shown formally: assuming the
infection is discovered at time zero, an integration of
eq.~(\ref{original_1}) yields $\n (T) = \n (0) \exp[\int_0^T
(\GR-c(\tau)) d\tau]= \n (0) \cdot e^{E(T)}$, where the effective
exponent

\begin{equation}
E(T)=\GR\cdot T - C(T)
\label{Eff_exponent}
\end{equation}

contains the accumulated medicine dose, $C(t)$, of
eq.~(\ref{integrated.medication}). Maximal suppression of the pathogen
population requires the largest achievable negative values of $E(T)$: the
positive growth term `$\GR\cdot T$' has to be minimized. This shows the
medication has to be given immediately.

The effective exponent also yields the condition, $E(T)=0$, which
estimates where the medication just balances pathogen growth. Assuming, as
above, that $C(T) \stackrel{}{\approx} C_\infty$, we find that for values of
$T$ surpassing

\begin{eqnarray}
T_{max}=\frac{N D_0}{\GR \CR}\; ,
\label{Tmax}
\end{eqnarray}

$E(T)$ becomes positive and pathogen growth is no longer kept in check. A fixed
total dosage $C_\infty$ thus implies a natural constraint on the total
treatment time beyond which drug overdilution renders treatments ineffective.

\subsection*{Fighting persistent pathogens: Pareto Front}

Transition into and out of the persistent state ($\down, \up > 0$)
allows pathogens to avoid the effects of medication and shortens the
effective maximal stalemate-time considerably, thus our estimate for
$T_{max}$, derived for the case of non-persisting pathogens, only
establishes an upper bound on the permissable total treatment time for
an effective treatment of persisters.

Due to their persistence ($\down, \up > 0$) pathogens show a delayed
response~\cite{Balaban04} (compare Fig.~\ref{fig1}) which
complicates their eradication. We now compare different eradication
strategies. First, values for~$D_0$ and $\CR$ (keeping the total
effective dose~$C_\infty$ constant), and a fixed total treatment
time $T$ are chosen. Then we vary the (ten) medication times
($\{t_k\}, k=1,...,10$) thus modifying the dosage strategies (choice
of time-points $t_k \in [0,T]$ in eq.~(\ref{Dosages.over.time})).
Upon integration of equations (\ref{original_1})
and~(\ref{original_2}), using the fourth order Runge-Kutta method,
we determine the number of survivors $\n(T)$ and $\p(T)$ as our
quality criterion.

The delayed response leads to a tradeoff between eradication of active versus
persister subpopulations, this complicates the analysis; without further
assumptions a best treatment strategy cannot be identified. To map out the
solution space, we therefore perform multi-objective
optimization~\cite{deb99Algorithm}, using
the NSGA-II genetic algorithm~\cite{multi.objective.optimization}. We
determine the set of Pareto-optimal strategies: in an $\n(T)$-over-$\p(T)$
plot they form a Pareto-optimal
front~\cite{deb99Algorithm} of points
corresponding to dosage strategies that lead to simultaneously minimized
(non-dominated) final values of $\n(T)$ and $\p(T)$.

For the integration of the differential equation, we use a regular
fourth-order Runge-Kutta integrator, with a step length of $\Delta t=0.01$.
The integration was found to provide consistent results for all runs up to a
step length of at least $\Delta t=0.015$, thus ensuring numerical stability of
the employed integration routine.

The NSGA-II multiobjective optimization
algorithm~\cite{multi.objective.optimization} with a population size
of 100 was used, running for 500 generations, using the final
populations $n(T)$ and $p(T)$ as the two objectives. The objectives
were constrained to nonnegative values to prevent the genetic
algorithm from being caught in spurious numerical instabilities. The
algorithm optimized the medication times $\{ t_k \}$ ($k=1\dots10$ in
our case), the crossover probability was set to~0.9 and the mutation
probability to~0.1. The parameters $\eta_c$ and $\eta_m$ for the
polynomial distributions used in the SBX crossover and in the mutation
operator~\cite{multi.objective.optimization,Deb95} were both set to
16.

The precise choice of these parameters turned out to be uncritical. We
found that the results reported below are robust with respect to
variations of crossover and mutation probabilities from $0.01$ to
0.99, similarly the $\eta_c$ and $\eta_m$ parameters could be varied
between 1.6 and 160 without affecting the position of the Pareto
front. Typically, the essential features of the Pareto front emerged
reproducibly after approximately 50 generations, whereas the remainder
of the run served to fine-tune the precise features of the front and
the corresponding solutions. Only for extreme choices of the
parameters, namely crossover and mutation probabilities equal to 0.01
and very narrow SBX characteristics, $\eta_c=160$, $\eta_m=160$, was
the extent of the Pareto front covered significantly more slowly;
apart from such extreme choices only insignificant performance
differences could be observed.

\begin{figure}
\begin{center}

\textbf{a}\\
\hspace{0.5cm}
\includegraphics[width=3.3in,height=2.5in]{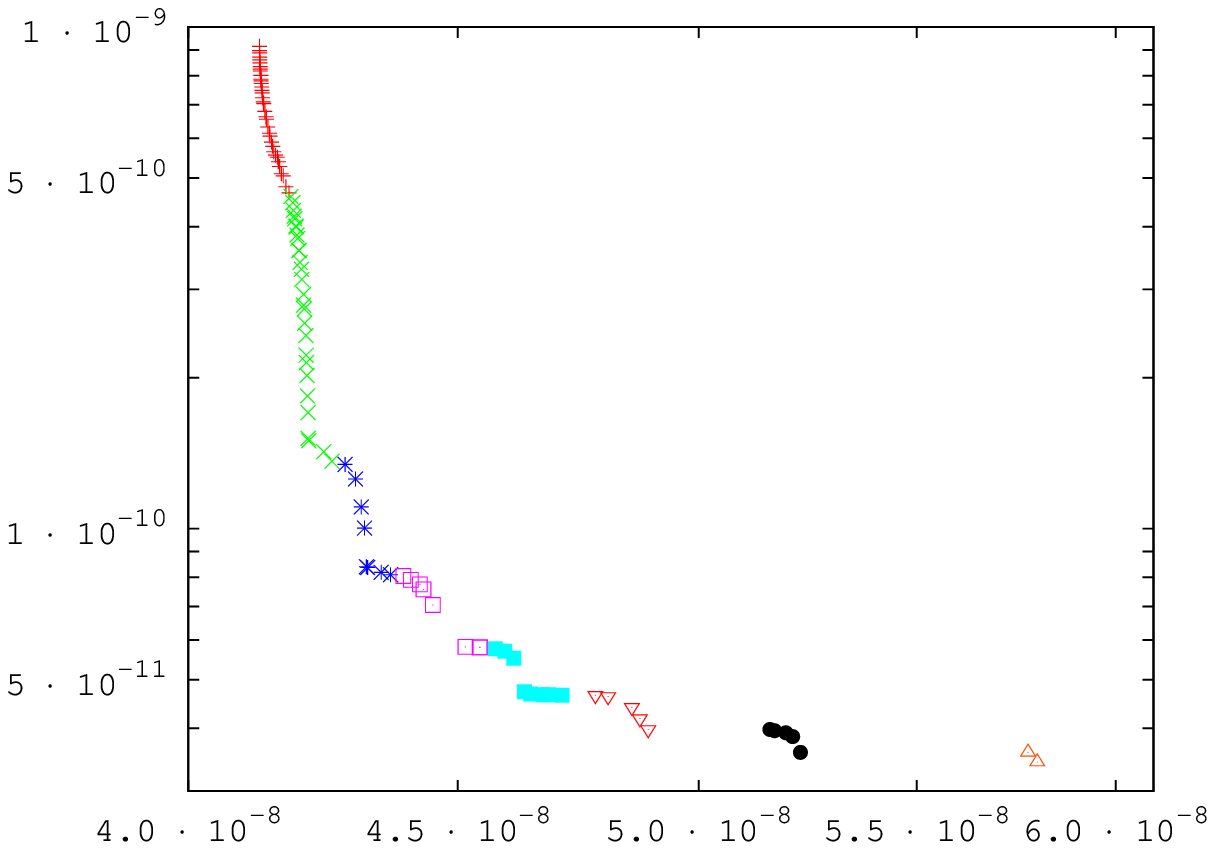}
\put(-245,20){\rotatebox{90}{\small relative number of active
bacteria~$\n(T)$}} \put(-180,-15){\small{relative number of
persisters~$\p(T)$}}
\\[4ex]
\textbf{b}\\[-1cm]
\hspace{0.5cm}

\raisebox{-0.5cm}{\includegraphics[width=3.75in,height=2.5in]{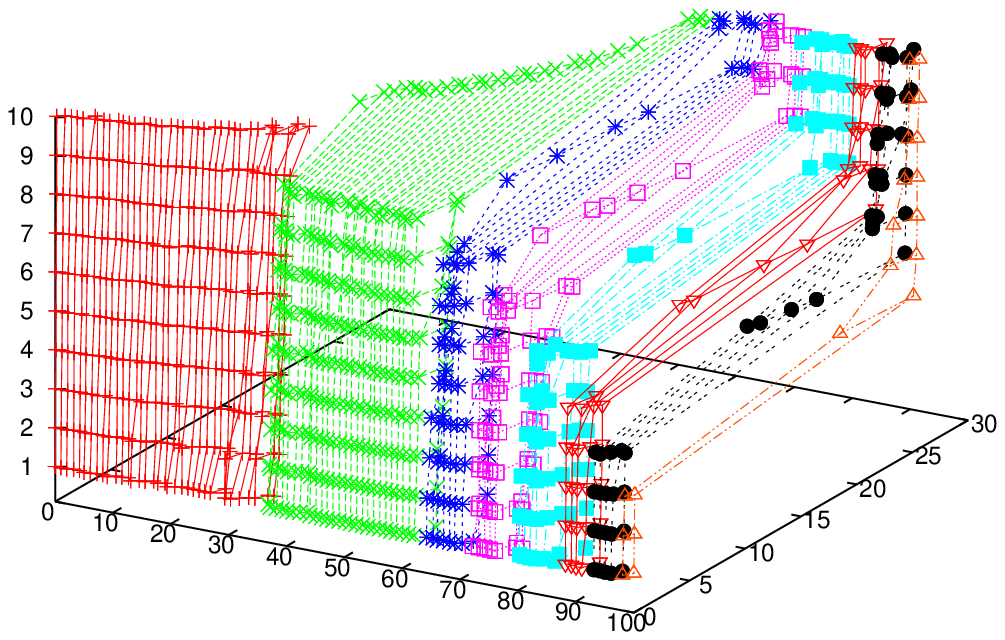}}
\put(-265,50){\rotatebox{90}{dose number $k$}}
\put(-210,10){\rotatebox{-9}{solution index $m$}}
\put(-70,0){\rotatebox{35}{times $t_k$ (h)}}

\end{center}

\caption{\sf \small Typical example of optimal treatment
  strategies.  Treatment time~$T=30 \,\mathrm{h}$, bacterial
  parameters of \Ecoli~ wildtype~\cite{KusselGenetics05} (except for
  our choice of $\mu_{\p}=0$): $\GR=2\,\mathrm{h}^{-1}, \down=1.2\cdot
  10^{-6}\,\mathrm{h}^{-1}, \up =0.1\,\mathrm{h}^{-1}$.  We choose
  $\n(0)=1$ and $\p(0)\approx5.714\cdot10^{-7}$ in accord with the
  natural initial condition~(\ref{init_ratio}). Medication parameters
  $D_0=100, N=10, \sigma=10 \,\mathrm{h}$ and
  $\CR=0.2\,\mathrm{h}^{-1}$. {\bf a}, The Pareto front of optimal
  strategies shows the tradeoff between suppressing active and
  persister subpopulations. The response margin for suppressing
  persisters is relatively much narrower than that for suppressing
  active pathogens (the spread of values on the horizontal coordinate
  axis is small).  {\bf b}, Same treatment regimes, plot displays dose
  number $k$ administered at time $t_k$ over total time~$t$ and
  index~$m$ (representing the solution number - out of 100) of
  solutions found by the NSGA-II algorithm sorted with respect to
  increasing values of surviving persisters~$\p(t)$. Note the
  emergence of distinct steps separating groups of treatment regimes
  despite a large value of $\sigma$ (this distinction becomes clearer
  still for smaller values of $\sigma$). The color- and symbol-coding
  matches subsections of the Pareto front in {\bf a} with the
  treatment regimes displayed in~{\bf b}.}
 \label{fig2}
\end{figure}

We now discuss the features of the solutions in detail:

At one end of the Pareto front one finds the strongest suppression of
persisters, at the other end the strongest suppression of active
bacteria, compare Fig.~\ref{fig2}~{\bf a}.

Fig.~\ref{fig2}~{\bf b} shows that a strategy aiming to suppress the
persistent subpopulation requires early administering of large doses
of medication. This is due to the fact that the active population has
to be suppressed early on and then some medication has to be used to
hold the persister's in check that are ``waking up'' and become active
again.  Alternatively, when the primary strategic aim is the
suppression of the active subpopulation or a mixed strategy, a later
application of the bulk of the medication is advised, although
(depending on details) some medication should also be given at the
start (as soon as the infection is discovered).  Associated optimal
strategies may therefore be very different from strategies which aim
for uniform constant exposure to medication, or, from the `kill before
they multiply' strategy described above. One should note the emergence
of discrete `bands' of optimal strategies in
Fig.~\ref{fig2}~\textbf{b}.
\begin{figure}
\begin{center}
\hspace{0.5cm}
\includegraphics[width=3.3in,height=2.4in]{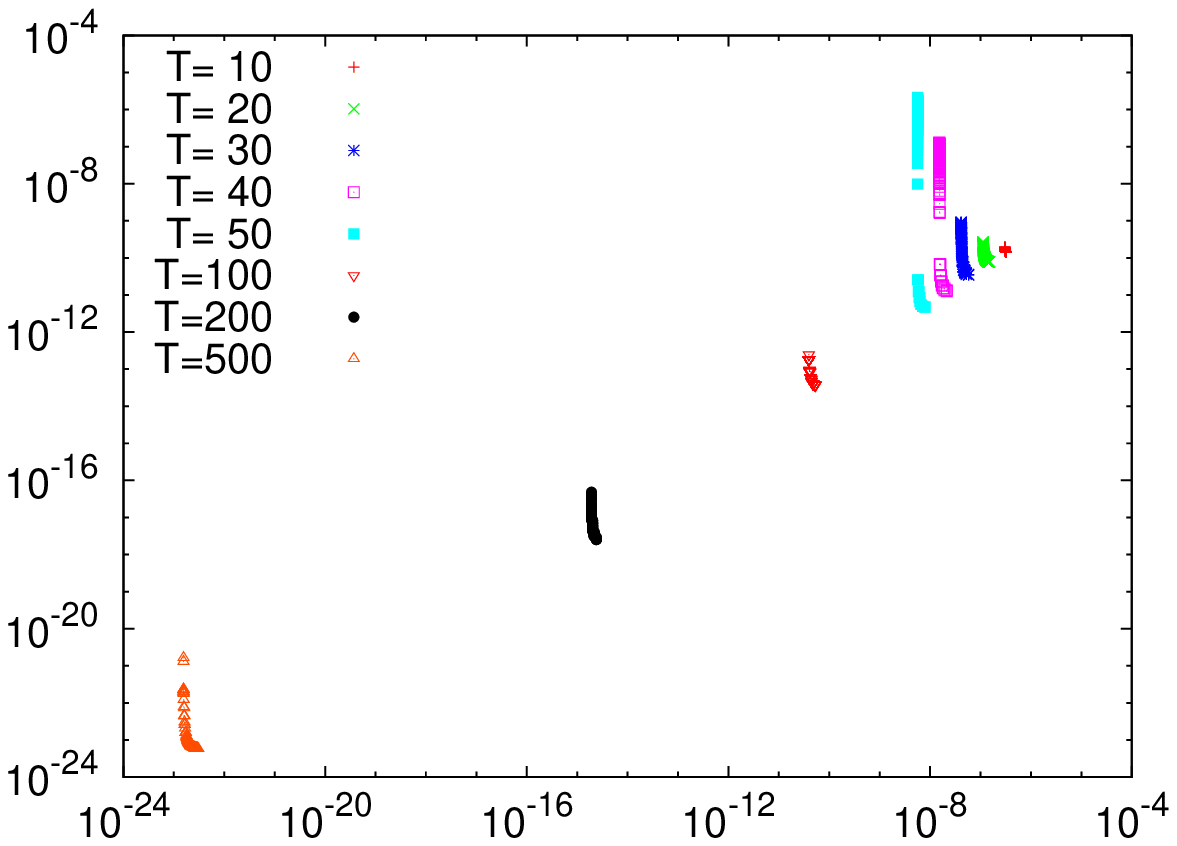}
\put(-240,10){\rotatebox{90}{\small relative number of active
pathogens~$\n(T)$}}
\put(-180,-15){\small{relative number of persisters~$\p(T)$}}
\end{center}
\caption{\sf \small A collection of several Pareto
fronts for various
  treatment times~$T$ shows increased effectiveness of longer
  treatments. Same parameters as in Figure~\ref{fig2} except for
  treatment times ranging from $T=10 \,\mathrm{h},\ldots,500
  \,\mathrm{h}$, see legend. We observe breakdown of treatment at
  $T\approx 600 \,\mathrm{h}$ (not shown) because medication becomes
  too much diluted.
\label{fig3}}
\end{figure}

Although the relative population suppression factor due to the medication
treatment may be satisfactory in the example sketched in
Fig.~\ref{fig2}~{\textbf{a}} and~\ref{fig3}, we are clearly most interested in
the critical cases where (because of cost, toxicity, or other reasons) the
treatment is in danger of failing. In this context it should be pointed out
that, worryingly, the narrow response margin of final persister
subpopulations~$\p(T)$ is a generic feature (see Fig.~\ref{fig2}). We are thus
led to consider a variation of the total treatment time~$T$ as well.
Fig.~\ref{fig3} displays several Pareto fronts for different values of~$T$.
Each individual front shares the features displayed in Fig.~\ref{fig2}.
Overall, there is a trend to more effective treatments with lengthened
treatment time in which case the medication has to be administered more
uniformly over the entire treatment period. It must not be lengthened too much
though, because the medication would become too diluted (see discussion
leading up to expression~(\ref{Tmax}) above).

\section*{Discussion}

The overall treatment time~$T$ has to be sufficiently long to kill persisters
which are protected by the time lag in the pathogens' response dynamics but
short enough not to dilute the medication concentration too much. If short
treatment times can yield sufficient pathogen suppression, the use of such
strategies may well be safer, since they lead us away from a possible breakdown
due to overdilution.

Also, as one shortens $T$, the characteristics of optimal strategies
change: instead of being uniformly distributed over $T$, doses are
typically increasingly concentrated at the beginning and/or end of the
treatment time, compare Fig.~\ref{fig2}~{\bf b}. Such strategies do
not only need less time than drawn out therapies, but they are also
simpler to administer. We believe that in view of widespread problems
with patient adherence to long lasting medication
regimens~\cite{Atreja05} such optimized strategies may offer relevant
alternatives that deviate from current clinical practice. In this
context we would like to point out that these treatment regimes appear
to be quite stable with respect to small changes in strategy, such as
completely concentrating all medication towards beginning and end of
the treatment period. In other words, a judiciously dosed two-shot
approach, almost the simplest conceivable strategy, can yield nearly
optimal results, is shorter than a maximally drawn out
therapy and not in danger of failure due to overdilution of the medication.

\begin{figure}
\begin{center}
\includegraphics[width=3.3in,height=2.4in]{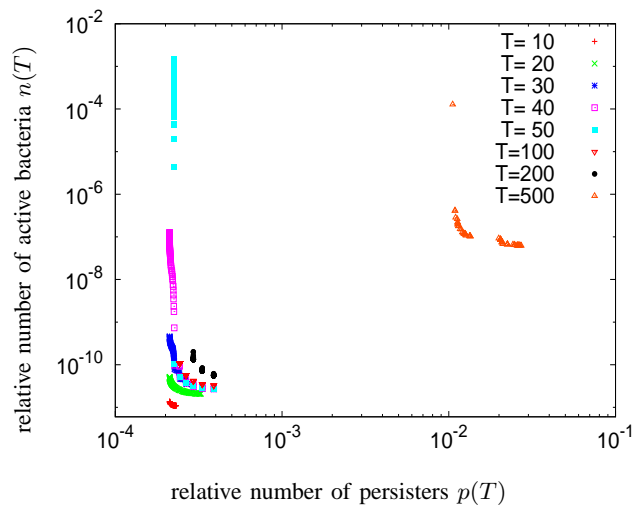}
\put(-240,10){\rotatebox{90}{\small relative number of active
bacteria~$\n(T)$}} \put(-180,-15){\small{relative number of
persisters~$\p(T)$}}
\end{center}
\caption{\sf \small Mutant pathogen infection is
incurable by the approach
  displayed in Figure~\ref{fig3}. A collection of several
  Pareto-fronts for various treatment times~$T$ (hours), see inset,
  using the same parameters as in Figures~\ref{fig2} and~\ref{fig3}.
  This includes use of the same initial conditions as in the wildtype
  (which are not the equilibrium conditions from~(\ref{init_ratio}) to
  allow for comparison); only $\down=10^{-3}\,\mathrm{h}^{-1}, \up
  =10^{-5}\,\mathrm{h}^{-1}$ are altered to describe the
  \emph{hipQ}-variant of \Ecoli~instead of its
  wildtype~\cite{KusselGenetics05}.
  \label{fig4}}
\end{figure}

For sufficiently large medication doses, the qualitative results
reported above also apply to the case of amplified persistence (as is
the case for the high persistence (\emph{hip}) mutants of
\Ecoli~analyzed in Balaban et al.~\cite{Balaban04}). When persistence
is increased, but a simultaneous increase in medication is impossible,
the results can change dramatically. An eradicable disease can become
unstoppable. To illustrate this point, we compare the response of the
wildtype of \Ecoli, Fig.~\ref{fig3}, with its highly persistent
\emph{hipQ}-type twin~\cite{Balaban04,KusselGenetics05} in
Fig.~\ref{fig4}. We assume the two to be identical except for their
different persistence rates $\down$ and
$\up$~\cite{KusselGenetics05}. Under identical treatment and initial
conditions as for the wildtype, we let the search algorithm find the
modified Pareto-front. Fig.~\ref{fig4} shows that short, highly
concentrated treatments allow us to suppress the active subpopulation
but they are too short to affect the persisters. For longer
treatments, the persisters' reflux rate $\up$, back to the normal
state, is still too low to deplete them sufficiently, the pathogen has
become untreatable.

%
%
%
When our simplifying assumption that persisters are entirely
resistant to medication is modified, in favour of reduced
susceptibility to medication, an extra term of the form `$- \omega
\cdot c(t) \cdot \p(t)$' has to be added to the right hand side of
eq.~(\ref{original_2}). With a reasonable factor of the order of
$\omega \approx \mu_\p / \GR$ ($ \approx 0.1$ in the case of
\Ecoli~\cite{KusselGenetics05}), our model still displays similar
generic features for optimal treatments. Regimes still form groups
of distinct strategies, doses for optimal treatment over
intermediate lengths~$T$ are still administered early and late. The
greatest difference is due to the greater vulnerability of the
bacteria (i.e., their smaller overall survival rates).

Finally, for other, different conditions and scenarios, such as
modified toxicity behaviour, nonlinear
dose-response~\cite{Boonkitticharoen90} and modified quality
criteria, one can also use multi-objective optimization to explore
and map the pertinent optimality regimes.

%

\paragraph*{Acknowledgments}

We thank K. Deb for providing us with the code for the NSGA-II
algorithm and helpful comments. Discussions, critical reading of the
manuscript and suggestions for improvement, by Femke van den Berg, Tim
Aldsworth, and Jan T. Kim, are gratefully acknow\-ledged.

\bibliographystyle{plain}
\bibliography{IEEE_MultiObj_Persister}

\end{document}